%% file: main.tex
\documentclass[conference]{IEEEtran}

\IEEEoverridecommandlockouts                              
\usepackage{graphicx}
\usepackage[utf8]{inputenc}
\usepackage[pdftex]{hyperref}
\usepackage{url}

\title{What Makes People Bond?: A Study on Social Interactions and Common Life Points on Facebook}

\author{
\IEEEauthorblockN{
Emanuel Sanchiz\IEEEauthorrefmark{1},
Francisco Ibarra\IEEEauthorrefmark{2},
Svetlana Nikitina\IEEEauthorrefmark{2}\IEEEauthorrefmark{3}, 
Marcos Báez\IEEEauthorrefmark{2} and
Fabio Casati\IEEEauthorrefmark{2}}

\IEEEauthorblockA{
\IEEEauthorrefmark{1}
Catholic University ``Nuestra Señora de la Asunción"\\
Tte. Cantaluppi y G. Molinas, Asunción, Paraguay\\ 
}
\IEEEauthorblockA{
\IEEEauthorrefmark{2}
University of Trento\\
Via Sommarive 9, Trento, Italy\\
}
\IEEEauthorblockA{
\IEEEauthorrefmark{3}
Tomsk Polytechnic University\\
Lenin Ave, 43, Tomsk, Russia \\
}
}

\begin{document}

\maketitle
\thispagestyle{empty}
\pagestyle{empty}

\begin{changemargin}{0cm}{0cm} 
\begin{abstract}
In this paper we aim at understanding if and how, by analysing people’s profile and historical data (such as data available on Facebook profiles and interactions, or collected explicitly) we can motivate two persons to interact and eventually create long-term bonds. We do this by exploring the relationship between connectedness, social interactions and common life points on Facebook. The results are of particular importance for the development of technology that aims at reducing social isolation for people with less chances to interact, such as older adults\footnote{
\copyright2016 IEEE. 
To be published in The 2016 International Conference on Collaboration Technologies and Systems (CTS 2016).

Personal use of this material is permitted.
Permission from IEEE must be obtained for all other uses, in any current or future media, including reprinting/republishing this material for advertising or promotional purposes, creating new collective works, for resale or redistribution to servers or lists, or reuse of any copyrighted component of this work in other works must be obtained from the IEEE.

For more details, see the \href{http://www.ieee.org/publications_standards/publications/rights/copyrightpolicy.html}{IEEE Copyright Policy.}

}. 
\end{abstract}

\end{changemargin}

{\keywords social interactions, homophily, older adults, empirical study}

\input{introduction}

\input{relatedworks}
\input{methods}

\input{results}

\input{discussion}
\section*{Acknowledgment}
This project has received funding from the EU Horizon
2020 research and innovation programme under the Marie Skłodowska-Curie grant agreement No 690962. This work was also
supported by the Trentino project ``Collegamenti" and by the project “Evaluation and enhancement of social, economic and emotional wellbeing of older adults”, agreement no. 14.Z50.310029, Tomsk Polytechnic University.

\bibliographystyle{IEEEtran}
\bibliography{references}

\end{document}

%% file: introduction.tex

\section{Introduction}

Being socially connected can have a significant impact on the quality of life of older adults. Research has demonstrated the association between health risks and the lack of social network diversity, infrequent contact with network members, and the small size of social networks \cite{brummett2001characteristics,seeman1994social,berkman1979social}. 

Social integration with peers is particularly important for older adults transitioning to residential care. Social integration helps in the adaptation, can foster friendships and sense of belonging, and has been found to be one of the key elements contributing to the quality of life in residential care \cite{bradshaw2012living}. Instead, failing to socially integrate contributes to feelings of loneliness, boredom, and helplessness, which are commonly regarded as the plagues of nursing home life \cite{thomas1996life}.

The research and practice on technology-supported social interactions in this context has mainly focused on \emph{enabling} social interactions (see, e.g., \cite{choi2012computer,cattan2005preventing} for a review), and less in addressing non-technological barriers, motivating social interactions and creating bonding. Addressing this gap requires the study and development of solutions that take into account the users’ needs, motivations and barriers. 

In our previous work \cite{baez2016personalized} we reported on the results from surveys and visits to nursing homes. We identified that
i) friendships in nursing homes are difficult, especially in the transition period, and that 
ii) contact is rather infrequent between older adults and their relatives, especially younger adults, often due to the lack of common topics of conversation and the lack of time. We suggested that technologies should go beyond \emph{enabling} interaction, to aim at creating 
friendships between people and opportunities for meaningful conversations. 

In this paper we follow up on these initial results and report on an exploratory study trying to understand the relationship between \emph{connectedness} among friends, \emph{social interactions} and \emph{common life points} on Facebook. 

The goal of this study is to understand if, by looking at information of the kind available in people’s Facebook profiles and posts, we can predict the feeling of connectedness between two Facebook friends and the intensity of their face-to-face interactions. Specifically, we investigate the following research questions:\\
\begin{itemize}
\item \textbf{RQ1. To what extent can we predict, by looking at profile information on Facebook, the frequency of online and offline communication between two persons?}
We are interested in understanding if common life points and social interactions are related, and whether certain common aspects can trigger interactions.

\item \textbf{RQ2. To what extent can we predict, by looking at profile information and intensity of social interactions, the feeling of connectedness between two persons?} This question is fundamental as it will help us understand whether having common aspects and a certain level of interaction is related to connectedness. Connectedness in this context represents the possibility of creating long-term bonds and friendship.
\end{itemize}

We explore the above questions in the broad population of Facebook users, from younger (18+) to  older adults (65+), since we are interested in \emph{intergenerational} as well as in \emph{peer} friend relationships.

In what follows we detail on the motivations, methods and results.



%% file: relatedworks.tex

\section{Background}

\subsection{Technologies to reduce social isolation}

Extensive work has been devoted to interventions aiming to reduce social isolation with the help of technology (e.g., \cite{choi2012computer,cattan2005preventing} for a review). Technology used to enable interactions for older adults include internet and email (e.g., \cite{blavzun2012impact}), social networks (e.g., \cite{ballantyne2010feel}), video chats (e.g., \cite{szeman2014new}), virtual companions (e.g., \cite{machesney2014gerontechnology}), and phone calls (e.g., \cite{cattan2011use}). Most one-to-one interventions limit the contact to a predefined person, such as a trained interviewer, a trained helper, or a volunteer \cite{machesney2014gerontechnology, cattan2011use}. However, interventions enabling social interactions with relatives and friends are more common in recent literature \cite{szeman2014new, ballantyne2010feel}. Interactions between participants and new people are also explored in some interventions \cite{fokkema2007escape}, in particular in those studying the effect of general internet use and social networks \cite{szeman2014new, blavzun2012impact}.

Research on online social interaction with older adults has focused more on ``enabling" communication and sharing, and less on creating opportunities for these interactions to happen. This calls for the development of technology that looks into making these interactions more effective. 

\subsection{Studies on friendship and common life points}

The notion that similarities among people lead to creating ties between them is known as homophily \cite{mcpherson2001birds}. In a review, McPherson et al. \cite{mcpherson2001birds} described it as \emph{``the principle that a contact between similar people occurs at a higher rate than among dissimilar people"}.

Homophily can be defined from two perspectives: i) \emph{value homophily}, which is based on the attitudes, beliefs and values, and ii) \emph{status homophily}, which is based on the major demographic dimensions such as race, ethnicity, sex, age, and characteristics like religion, education, occupation \cite{mcpherson2001birds}, \cite{lazarsfeld1954friendship}. 
A review of studies done by Fehr \cite{fehr2008friendship} suggests that both status and value homophily are relevant for building friendship. However, a recent survey by Campbell shows that only value homophily affects friendship chemistry (emotional and psychological connection between persons) \cite{campbell2015friendship}. 

There are studies analysing structural properties of friend networks \cite{traud2012social} and empirical studies that have explored homophily in social networks. Kwak \cite{kwak2010twitter} studied homophily among Twitter users (with 1000 and less followers) and their friends-followers and found the effect for geographic location and popularity. Lewis et al. \cite{lewis2008tastes} studied Facebook profiles of 1640 college students in the US and found significant shared interests (movies, music, books) for certain connections (being Facebook friends, picture friends and reciprocal tagging). A similar study by Bobo et al. \cite{nick2013simmelian} analysed a Facebook dataset of 100 US Universities and concluded that homophily by dormitory, graduation year, and gender is strong.

The above ideas have also been applied to algorithms. In the literature, the approaches used to match friends can be generally classified as content-based and link-based.

Algorithms relying on content, use the similarity of users' profiles in order to make friend recommendations. This implies comparing what users state in their profiles to keywords and tags from other profiles  \cite{chen2009make}. This general approach has been successfully used for recommending books, movies and web sites (e.g., \cite{mooney2000content}). Link-based algorithms (e.g., friend-of-friend) use social network information only,  relying on the idea that if two persons have a lot of friends in common, perhaps they could be friends. For example, the Facebook feature ``people you may know" is partially based on this approach \cite{chen2009make}.

In this work, we build on the notion of homophily - which has been largely studied - but unlike previous works we focus on predicting the feeling of connectedness and social interactions. Our results could inform approaches for recommending friends and conversation topics.

%% file: methods.tex
\section{Methods}
\subsection{Hypotheses}

In this exploratory study, we specifically investigate the following hypotheses:\\



\noindent\textbf{H1. Common life points are related to the level of online and face-to-face interactions}. 
This will help us understand if and how we can predict the frequency of social interactions based on the similarity of people (RQ1).

\noindent\textbf{H2. Connectedness is related to common life points and both online and face-to-face interactions}. 
This will tell us if and how we can predict connectedness on the basis of similarity of users and their frequency of interaction (RQ2). 

We should notice that the above corresponds to a preliminary work, in which we are setting the direction for further analysis. We do not assume any causal relationship, which should be tested with a controlled trial.

\begin{figure*}
\centering
\includegraphics[width=\textwidth]{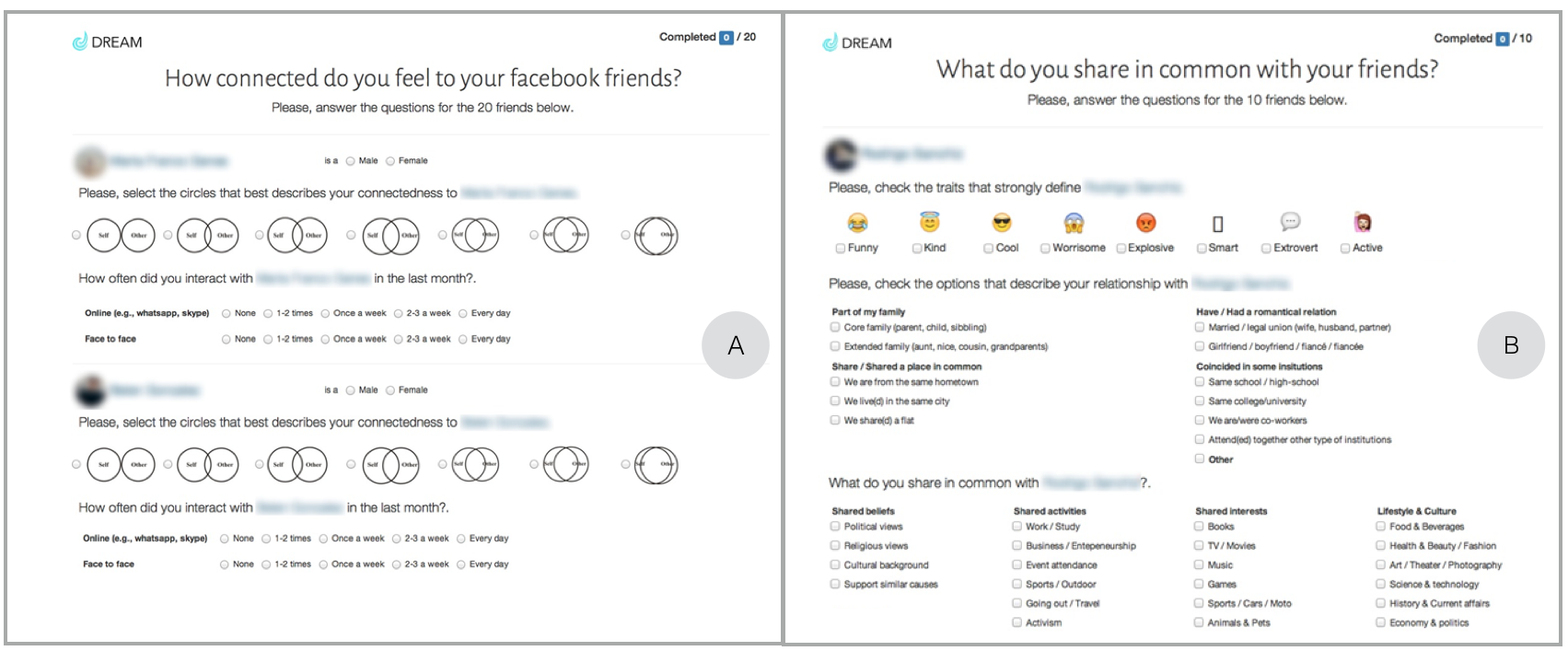}
\caption{FriendRover application. a) Connectedness form, and b) common life points form. }
\label{figure:friendrover}
\end{figure*}

\subsection{Data collection}

We collected information from Facebook users, both automatically (from users' profile, with users' permission) and by explicitly  asking users about the frequency and nature of their interactions with friends, as well as the level of connectedness they feel with friends. We analysed profile information (specifically the common aspects between people’s profiles) and interactions to build a model for predicting connectedness and actual face-to-face interactions. In other words, our variables are:

\begin{itemize}
\item \textbf{Connectedness}. Measured using an adaptation of the Inclusion of Other in Self (IOS) scale by Aron et al. \cite{aron1992inclusion}, a 7-point scale that relies on pictograms.
\item \textbf{Social interactions}. Described in terms of \emph{online interactions} and \emph{face-to-face interactions}, both measured on a 5-point frequency scale.
\item \textbf{Common life points}. Described in terms of \emph{shared relationships} (family ties, having lived in the same places, having attended the same institutions), and \emph{shared aspects} (shared beliefs, 
activities, and 
interests).
\end{itemize}

To collect the information needed for the analysis we developed a Facebook application called FriendRover\footnote{Available at: \url{http://happy.mateine.org/friends}}. 
The workflow of the application is illustrated in Figure \ref{figure:friendrover} and detailed below:
\begin{itemize}
\item  Users open the application and are presented with instructions as well as the request for consent. After giving consent and logging in, the data on participants' Facebook profile, friends, posts, and interactions on their posts, is automatically collected and anonymised.
\item Then, to each participant we show a list of 20 friends who have interacted with the participant’s posts (through reactions, comments, and tags). These friends are selected in a way such that they are representative of different levels of interaction (We categorised friends into quartiles according to the interaction with the participant and then took a sample from each quartile). From this list, users report on \emph{connectedness} and \emph{social interactions} (Figure \ref{figure:friendrover} A).
\item Then we take the 10 friends rated as more connected by the participant, and for each friend we ask the participant to specify the traits that better describe this friend. On this interface participants report on \emph{common life points} (Figure \ref{figure:friendrover} B).
\end{itemize}


\subsection{Participants}
The study was conducted online with a convenience sample of Facebook users (over 18 years old), obtained by advertising the survey on the Facebook pages of members from the research team. Participants were eligible if they have interacted with at least 20 persons. 
For this study, we advertised the experiment among Spanish-speaking users.

\subsection{Resulting dataset}
We collected the responses of 33 participants (age range: 32-65, mean: 33 years old, 45\% female), which resulted in 660 friendship relationships. The dataset consists of 660 connectedness samples and 280 reports on common life points out of 330 possible reports, this is because some participants did not complete the second part in full.

%% file: results.tex

\section{Results}

\subsection{Common life points are related to the level of online and face-to-face interactions}
We addressed H1 by testing the association of common life points with online and face-to-face interaction separately.

An analysis of variance was performed to determine a statistically significant difference in the level of \emph{online interactions} for the number of common life points, using the number of \emph{shared aspects} and \emph{shared relations} as independent variables. The results show a main effect for number of shared aspects (F(1, 279)=57.268, p$<$.001) and a main effect for number of  shared relations (F(1, 279)=15.251, p$<$.001), but no interaction effect between both variables.

Analysing the individual components of both dependent variables we see a main effect for
\emph{shared activities} (F(1, 279)=27.535, p$<$.001) and
\emph{shared interests} (F(1, 279)=31.439, p$<$.001) but no main effect for shared beliefs (F(1,279)=0.996, p=.319). We also observe main effects for
\emph{common institution} 
(F(1, 279)=9.483, p=.002) and
\emph{common place}
(F(1, 279)=6.798, p=.01) but no main effect for \emph{family ties}
(F(1, 279)=1.547, p=.214).

\begin{figure}
\centering
\includegraphics[width=\columnwidth]{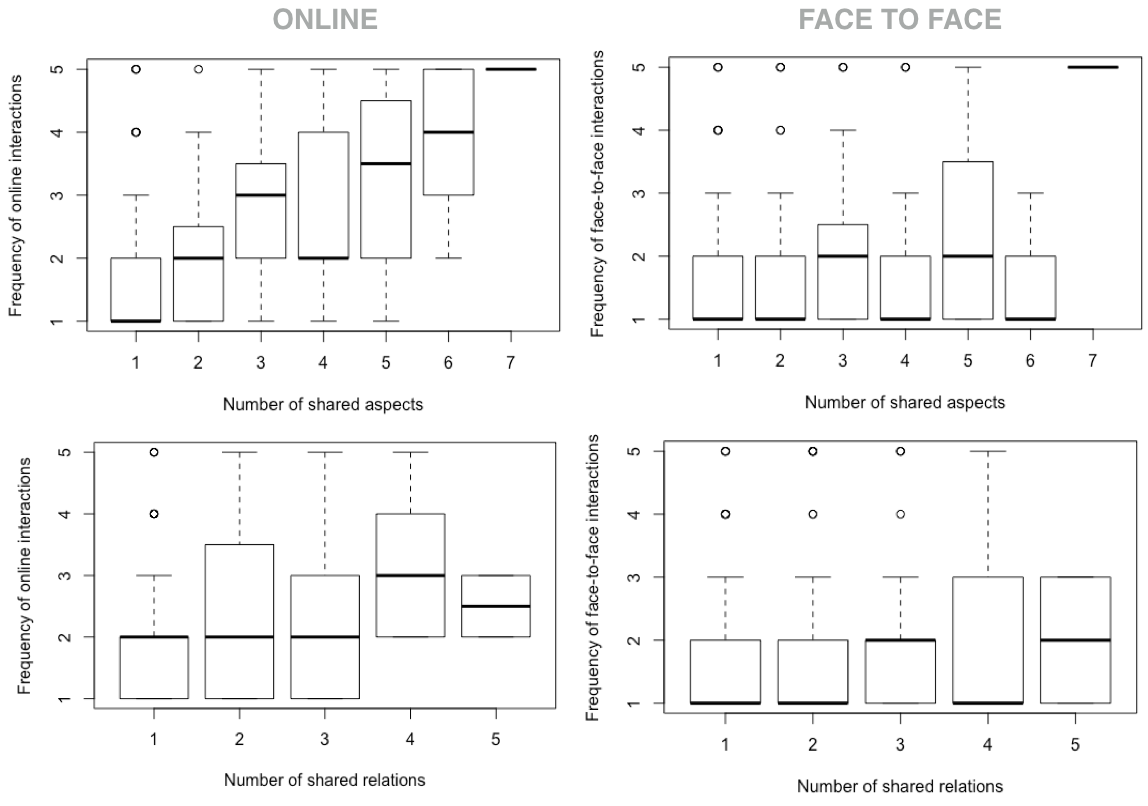}
\caption{ Common aspects and social interactions}
\label{figure:chart-h1}
\end{figure}

The above suggests that shared beliefs (religion, politics, cultural background, and causes) do not significantly help predicting online interactions when controlling for other factors. Likewise, social interactions do not significantly differ for relatives vs. non-relatives (family ties), when other factors are considered. Overall, as seen in Figure \ref{figure:chart-h1}, the relationship suggests that \textbf{the more aspects one shares, the more frequent the online interactions are} -- especially when there are common interests and when people engage in joint activities. This trend is not present in shared relations, where some relationships might be dominating the effect.

Using the same model with the level of \emph{face-to-face interactions} as dependent variable, the results show a main effect for number of shared relations (F(1, 279)=8.328, p=.004), a main effect for number of shared aspects (F(1, 279)=12.587, p$<$.001), and an interaction effect between both variables (F(1, 279)=9.420, p $=$.002). 

Replacing the independent variables for their individual components in the model, we see a main effect for 
\emph{shared activities} (F(1, 279)=5.388, p=.02) and
\emph{shared interests} (F(1, 279)=11.480, p$<$.001) but no main effect for shared beliefs. We also observe main effects for 
\emph{family ties} (F(1, 279)= 4.940, p=.027) and 
\emph{common locations} (F(1, 279)= 4.513, p=.034) but not for \emph{common institutions}. These results are similar to those for online interactions with the difference that \textbf{family ties become a relevant predictor of face-to-face interactions}.

\subsection{Connectedness is related to common life points and face-to-face interactions}

To test whether there is a significant difference in connectedness for the various levels of social interactions, we performed an analysis of variance with \emph{connectedness} as a dependent variable and the levels of \emph{online} and \emph{face-to-face} interactions as 
independent variables.

The results show a significant main effect for the level of face-to-face interactions (F(1, 659)=388.4, p$<$.001) and online interactions (F(1, 659)=218.5, p$<$.001), and also a significant interaction effect between both variables (F(1, 659)= 57.8, p$<$.001).

\begin{figure}
\centering
\includegraphics[width=\columnwidth]{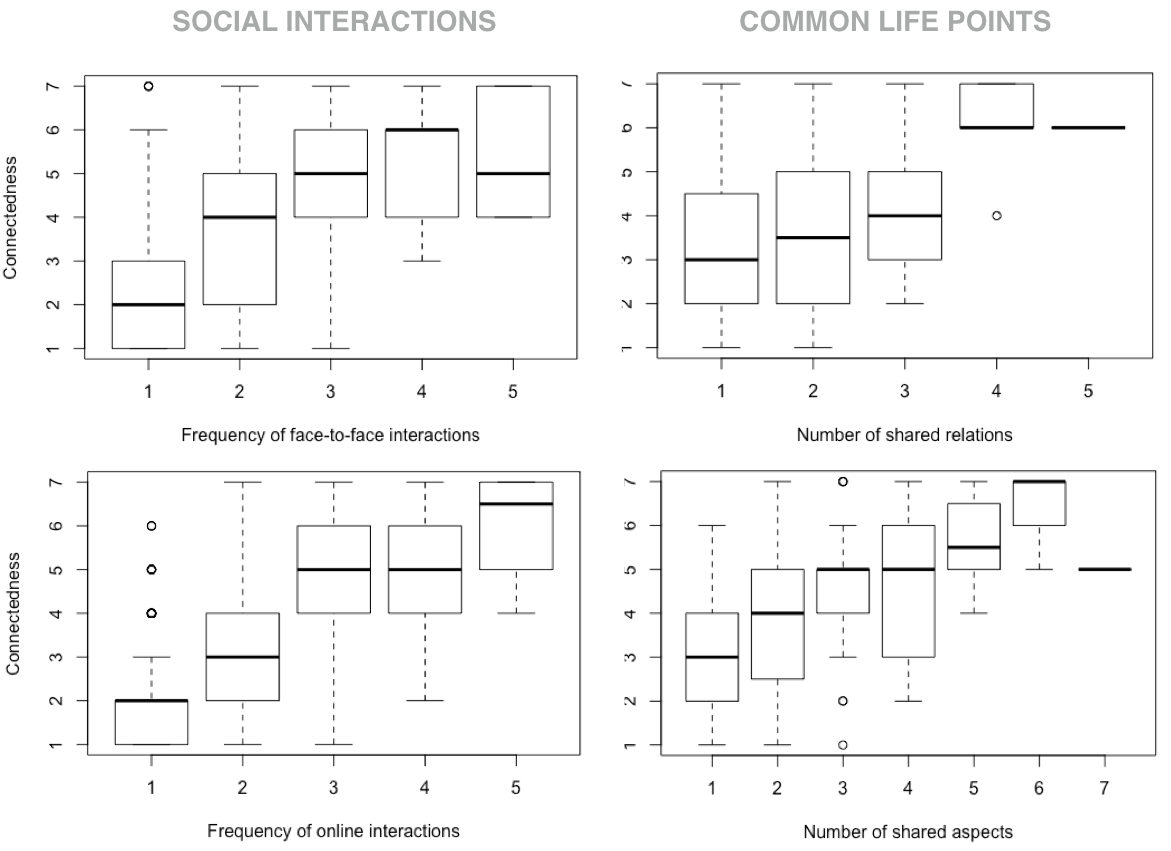}
\caption{ Common life points and connectedness}
\label{figure:chart-h2}
\end{figure}

We illustrate the above relationships in Figure \ref{figure:chart-h2}. For social interactions, the relation suggests a \textbf{higher level of connectedness for people interacting more frequently}. The outliers for the lowest levels of interaction correspond to people living abroad but interacting online very frequently (online), as well as people spending time together but not so much of this time online (face-to-face). This is an example of the interaction effect between both variables.
\begin{figure}
\centering
\includegraphics[width=\columnwidth]{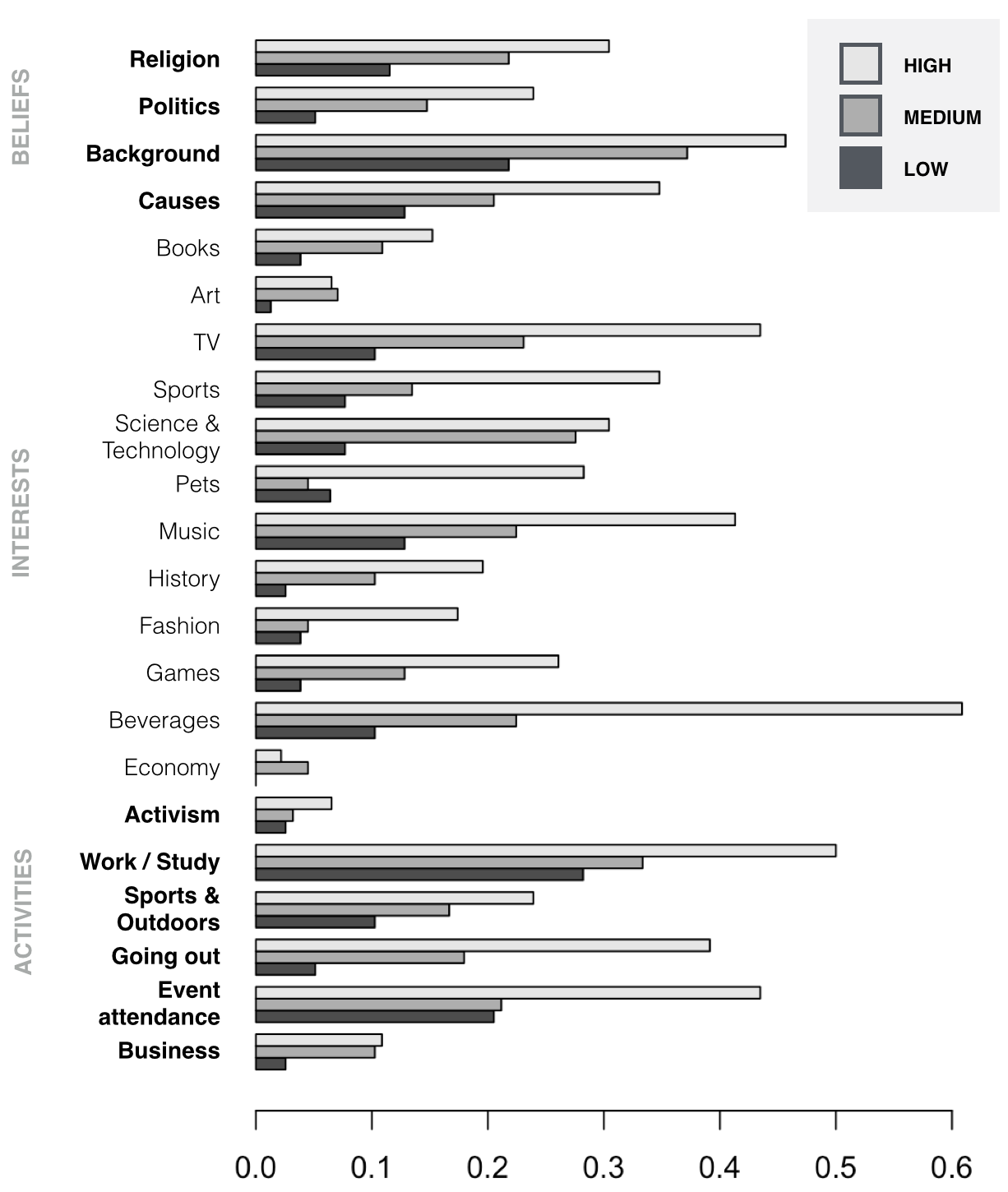}
\caption{Percentage of relationships featuring each common aspect, grouped and normalized by high, medium and low levels of connectedness }
\label{figure:chart-common-aspects}
\end{figure}

Analysing the relationship with common life points we observe main effects for the number of \emph{shared relationships} (F(1, 279)=43.48, p$<$.001) and for the number of \emph{shared aspects} (F(1, 279)=46.06, p$<$.001), but no interaction effect between both variables. These relationships are illustrated in Figure \ref{figure:chart-h2} and suggest that \textbf{having more common life points contributes to a higher level of connectedness}. 

More details are presented in Figure \ref{figure:chart-common-aspects}, showing the percentage of shared aspects by connectedness level. In the figure we can see a higher percentage of participants reporting sharing common aspects for higher levels of connectedness. The difference is more pronounced for shared interests.

%% file: discussion.tex

\section{Discussion}
In this paper we have explored the relationship between connectedness, social interactions, and common life points on Facebook in two main research questions. As an exploratory work, the questions were approached from a general perspective but still bringing some interesting insights. 

With respect to the relation between common life points and social interactions, we have seen that the more \emph{common life points} friends share the more frequent their \emph{online social interactions} are, and that shared interests and activities are determinant to this effect. For \emph{face-to-face social interactions} the relationship is more complex, with family ties becoming a relevant predictor. 

Interestingly, by exploring common life points we have seen that shared beliefs, as reported by the participants, is not a good predictor of social interactions, even when the literature points to this as a determinant factor \cite{mcpherson2001birds}. We argue that this might be due to the homogeneity of the participants targeted by the study (Spanish-speaking), or simply the limitation in the type of metadata available on Facebook. Moreover, we have seen that shared activities are strong predictors, which is in line with previous literature stating that accomplishing practical activities together strengthen social ties. 

We have also seen that higher levels of interaction and common life points are related to higher levels of connectedness. This suggests that one potential direction to creating bonds is generating opportunities for similar people to have meaningful interactions.

The above gives empirical support to technology aiming at increasing social interactions and creating long term bonds, by - for example - i) seeking to match users based on common life points, ii) generating conversations around shared interests, and iii) engaging users in shared activities. 


As for ongoing and future work, we plan to follow up on this study to extend it to 6 countries (Mongolia, Italy, Paraguay, Costa Rica, Russia, Philippines) and analyse cross-cultural as well as age-group differences. 
From a technological standpoint, we are currently incorporating these findings in the design of tools to reduce social isolation in older adults, including virtual environments with the dual purpose of performing productive activities (crowdsourcing / volunteering) and socializing online. The latter comes from the fact that social interactions are of particular importance when providing productive activities to older adults \cite{ibarratools}, and it is one example of how the findings of this paper can be applied to collaborative systems.





%% file: main.bbl
\begin{thebibliography}{10}
\providecommand{\url}[1]{#1}
\csname url@samestyle\endcsname
\providecommand{\newblock}{\relax}
\providecommand{\bibinfo}[2]{#2}
\providecommand{\BIBentrySTDinterwordspacing}{\spaceskip=0pt\relax}
\providecommand{\BIBentryALTinterwordstretchfactor}{4}
\providecommand{\BIBentryALTinterwordspacing}{\spaceskip=\fontdimen2\font plus
\BIBentryALTinterwordstretchfactor\fontdimen3\font minus
  \fontdimen4\font\relax}
\providecommand{\BIBforeignlanguage}[2]{{%
\expandafter\ifx\csname l@#1\endcsname\relax
\typeout{** WARNING: IEEEtran.bst: No hyphenation pattern has been}%
\typeout{** loaded for the language `#1'. Using the pattern for}%
\typeout{** the default language instead.}%
\else
\language=\csname l@#1\endcsname
\fi
#2}}
\providecommand{\BIBdecl}{\relax}
\BIBdecl

\bibitem{brummett2001characteristics}
B.~H. Brummett, J.~C. Barefoot, I.~C. Siegler, N.~E. Clapp-Channing, B.~L.
  Lytle, H.~B. Bosworth, R.~B. Williams~Jr, and D.~B. Mark, ``Characteristics
  of socially isolated patients with coronary artery disease who are at
  elevated risk for mortality,'' \emph{Psychosomatic Medicine}, vol.~63, no.~2,
  pp. 267--272, 2001.

\bibitem{seeman1994social}
T.~E. Seeman, L.~F. Berkman, D.~Blazer, and J.~W. Rowe, ``Social ties and
  support and neuroendocrine function: the macarthur studies of successful
  aging.'' \emph{Annals of Behavioral Medicine}, 1994.

\bibitem{berkman1979social}
L.~F. Berkman and S.~L. Syme, ``Social networks, host resistance, and
  mortality: a nine-year follow-up study of alameda county residents,''
  \emph{American journal of Epidemiology}, vol. 109, no.~2, pp. 186--204, 1979.

\bibitem{bradshaw2012living}
S.~A. Bradshaw, E.~D. Playford, and A.~Riazi, ``Living well in care homes: a
  systematic review of qualitative studies,'' \emph{Age and ageing}, p. afs069,
  2012.

\bibitem{thomas1996life}
W.~H. Thomas, \emph{Life worth living: How someone you love can still enjoy
  life in a nursing home: The Eden Alternative in action}.\hskip 1em plus 0.5em
  minus 0.4em\relax Publisher: VanderWyk\&Burnham, 1996.

\bibitem{choi2012computer}
M.~Choi, S.~Kong, and D.~Jung, ``Computer and internet interventions for
  loneliness and depression in older adults: a meta-analysis,''
  \emph{Healthcare informatics research}, vol.~18, no.~3, pp. 191--198, 2012.

\bibitem{cattan2005preventing}
M.~Cattan, M.~White, J.~Bond, and A.~Learmouth, ``Preventing social isolation
  and loneliness among older people: a systematic review of health promotion
  interventions,'' \emph{Ageing and society}, vol.~25, no.~01, pp. 41--67,
  2005.

\bibitem{baez2016personalized}
M.~Baez, C.~Dalpiaz, F.~Hoxha, A.~Tovo, V.~Caforio, and F.~Casati,
  ``Personalized persuasion for social interactions in nursing homes,''
  \emph{arXiv preprint arXiv:1603.03349}, 2016.

\bibitem{blavzun2012impact}
H.~Bla{\v{z}}un, K.~Saranto, and S.~Rissanen, ``Impact of computer training
  courses on reduction of loneliness of older people in finland and slovenia,''
  \emph{Computers in Human Behavior}, vol.~28, no.~4, pp. 1202--1212, 2012.

\bibitem{ballantyne2010feel}
A.~Ballantyne, L.~Trenwith, S.~Zubrinich, and M.~Corlis, ``'i feel less
  lonely': what older people say about participating in a social networking
  website,'' \emph{Quality in Ageing and Older Adults}, vol.~11, no.~3, pp.
  25--35, 2010.

\bibitem{szeman2014new}
Z.~Sz{\'e}man, ``A new pattern in long term care in hungary: Skype and youth
  volunteers,'' \emph{Anthropological notebooks}, pp. 105--117, 2014.

\bibitem{machesney2014gerontechnology}
D.~Machesney, S.~S. Wexler, T.~Chen, and J.~F. Coppola, ``Gerontechnology
  companion: Virutal pets for dementia patients,'' in \emph{IEEE LISAT
  2014}.\hskip 1em plus 0.5em minus 0.4em\relax IEEE, 2014, pp. 1--3.

\bibitem{cattan2011use}
M.~Cattan, N.~Kime, and A.-M. Bagnall, ``The use of telephone befriending in
  low level support for socially isolated older people--an evaluation,''
  \emph{Health \& social care in the community}, vol.~19, no.~2, pp. 198--206,
  2011.

\bibitem{fokkema2007escape}
T.~Fokkema and K.~Knipscheer, ``Escape loneliness by going digital: A
  quantitative and qualitative evaluation of a dutch experiment in using ect to
  overcome loneliness among older adults,'' 2007.

\bibitem{mcpherson2001birds}
M.~McPherson, L.~Smith-Lovin, and J.~M. Cook, ``Birds of a feather: Homophily
  in social networks,'' \emph{Annual review of sociology}, pp. 415--444, 2001.

\bibitem{lazarsfeld1954friendship}
P.~F. Lazarsfeld, R.~K. Merton \emph{et~al.}, ``Friendship as a social process:
  A substantive and methodological analysis,'' \emph{Freedom and control in
  modern society}, vol.~18, no.~1, pp. 18--66, 1954.

\bibitem{fehr2008friendship}
B.~Fehr, ``Friendship formation,'' \emph{Handbook of relationship initiation},
  pp. 29--54, 2008.

\bibitem{campbell2015friendship}
K.~Campbell, N.~Holderness, and M.~Riggs, ``Friendship chemistry: An
  examination of underlying factors,'' \emph{The Social science journal},
  vol.~52, no.~2, pp. 239--247, 2015.

\bibitem{traud2012social}
A.~L. Traud, P.~J. Mucha, and M.~A. Porter, ``Social structure of facebook
  networks,'' \emph{Physica A: Statistical Mechanics and its Applications},
  vol. 391, no.~16, pp. 4165--4180, 2012.

\bibitem{kwak2010twitter}
H.~Kwak, C.~Lee, H.~Park, and S.~Moon, ``What is twitter, a social network or a
  news media?'' in \emph{Proceedings of the 19th international conference on
  World wide web}.\hskip 1em plus 0.5em minus 0.4em\relax ACM, 2010, pp.
  591--600.

\bibitem{lewis2008tastes}
K.~Lewis, J.~Kaufman, M.~Gonzalez, A.~Wimmer, and N.~Christakis, ``Tastes,
  ties, and time: A new social network dataset using facebook. com,''
  \emph{Social networks}, vol.~30, no.~4, pp. 330--342, 2008.

\bibitem{nick2013simmelian}
B.~Nick, C.~Lee, P.~Cunningham, and U.~Brandes, ``Simmelian backbones:
  amplifying hidden homophily in facebook networks,'' in \emph{IEEE/ACM ASONAM
  2013}.\hskip 1em plus 0.5em minus 0.4em\relax IEEE, 2013, pp. 525--532.

\bibitem{chen2009make}
J.~Chen, W.~Geyer, C.~Dugan, M.~Muller, and I.~Guy, ``Make new friends, but
  keep the old: recommending people on social networking sites,'' in
  \emph{Proceedings of the SIGCHI Conference on Human Factors in Computing
  Systems}.\hskip 1em plus 0.5em minus 0.4em\relax ACM, 2009, pp. 201--210.

\bibitem{mooney2000content}
R.~J. Mooney and L.~Roy, ``Content-based book recommending using learning for
  text categorization,'' in \emph{Proceedings of the fifth ACM conference on
  Digital libraries}.\hskip 1em plus 0.5em minus 0.4em\relax ACM, 2000, pp.
  195--204.

\bibitem{aron1992inclusion}
A.~Aron, E.~N. Aron, and D.~Smollan, ``Inclusion of other in the self scale and
  the structure of interpersonal closeness.'' \emph{Journal of personality and
  social psychology}, vol.~63, no.~4, p. 596, 1992.

\bibitem{ibarratools}
F.~Ibarra, O.~Korovina, M.~Baez, G.~Barysheva, M.~Marchese, L.~Cernuzzi, and
  F.~Casati, ``Tools enabling online contributions by older adults,''
  \emph{IEEE Internet Computing}, vol.~PP, no.~99, pp. 1--1, 2016.

\end{thebibliography}
